**Tribochemistry, mechanical alloying, mechanochemistry: what is in a name?**

*Adam A. L. Michalchuk,[1] Elena V. Boldyreva,[2,3] Ana M. Belenguer,[4] Franziska Emmerling,[1] Vladimir V. Boldyrev[2,5]*

1. Federal Institute for Materials Research and Testing (BAM), Richard-Willstaetter Str 11, 12489, Berlin, Germany
2. Novosibirsk State University, ul. Pirogova, 2, Novosibirsk, 630090, Russian Federation
3. Boreskov Institute of Catalysis SB RAS, pr. Lavrentieva, 5, Novosibirsk, 630090, Russian Federation
4. Yusef Hamied Department of Chemistry, University of Cambridge, CB2 1EW, UK
5. Voevodski Institute of Chemical Kinetics and Combustion SB RAS, ul. Institutskaia, 3, Novosibirsk, 630090, Russian Federation

**ABSTRACT**

Over the decades, the application of mechanical force to influence chemical reactions has been called by various names: *mechanochemistry, tribochemistry, mechanical alloying,* to name but a few. The evolution of these terms has largely mirrored the understanding of the field. But what is meant by these terms, why have they evolved, and does it really matter how a process is called? Which parameters should be defined to describe unambiguously the experimental conditions such that others can reproduce the results, or to allow a meaningful comparison between processes explored under different conditions? Can the information on the process be encoded in a clear, concise, and self-explanatory way? We address these questions in this *Opinion* contribution, which we hope will spark timely and constructive discussion across the international mechanochemistry community.

## 1. INTRODUCTION

Chemical transformations initiated by mechanical energy appear to be the first reactions that humans learned to induce and control, even before thermal reactions were possible. In fact, the first combustion reactions were *mechano-chemical* (*tribo-chemical*), wherein fire was produced through friction. Throughout human history, mechanically induced chemical reactions have accompanied many significant technological advances. For example, since the discovery of black powder in *ca.* 220 BCE, explosives have allowed the advent of mining and have facilitated the construction of cities and infrastructures. More recently, the continued development of mechano-chemistry promises to revolutionize the chemical industry, providing synthetic routes devoid of environmentally harmful solvents.[1,2] The potential for mechanochemistry to have paradigm-changing impact across the chemical sciences has placed the field amongst IUPAC's '10 chemical innovations that will change our world'.[3]

The earliest written record of a mechano-chemical transformation seems to be that by Theophrastus of Eresus, in his book "On Stones" of *ca.* 315 B.C.[4] Theophrastus describes the reduction of cinnabar to mercury through grinding using a copper mortar and pestle. Although grinding and milling were used extensively over the centuries for the processing of grains, minerals, and even pharmaceuticals, mention of mechano-chemical processes in the scientific literature did not reappear until the 19$^{th}$ century. These early reports include those by Faraday (1820) on the dehydration of crystal hydrates,[5,6] Carey-Lee (1866) on the decomposition of silver, gold, and mercury halides on grinding,[7,8] and by both Ling and Baker (1893)[5,6] and Flavitsky[9,10] who described organic chemical reactions upon grinding. The attention of mechano-chemical investigation soon expanded to a wide range of material types, and explored an array of phenomena including the initiation of explosives by impact and friction,[11–14] and the mechanical-decomposition of polymers.[15–19] Similarly, mechano-chemical investigation into areas including the chemical processes accompanying mining, metallurgy, and the manufacturing of various oxide and chalcogenide materials became a prominent direction of research,[20–30] expanding towards the preparation and processing of fine chemicals and pharmaceuticals.[31–37] The 20th century represents a period of remarkable development of the fundamental aspects of mechano-chemistry and of significant progress towards scaling mechano-chemical reactions towards real-world industrial applications. Although progress in mechano-chemistry through the 20th century was dominated by studies of metals, inorganic compounds, materials, and catalysts, significant advances were also made in the mechano-chemistry of organic polymers and drug compounds and formulations.[20–25,33,33,38–41] A number of dedicated texts on the historical development of mechanochemistry are available elsewhere.[5,42,43]

To date, mechano-chemical approaches being applied to transformations from across the chemical sciences have been reported, spanning from the synthesis of organic compounds (e.g. peptides[44,45]) through to the preparation of large porous frameworks such as metal-organic frameworks (MOFs[46–51]). Moreover, the scale of mechano-chemical reactions has ranged from the mechanical manipulation of single atoms and molecules (predominantly, synthetic and natural polymers) using atomic force microscopy[52–55] to the induction of reactions in multi-component powder mixtures in ball milling reactors or extruders.[56–60] Alongside synthetic covalent chemical reactions, a wide range of supramolecular assemblies have been also prepared by mechanical treatment, including cocrystals and salts,[61–66] as well as non-covalently bound mechano-composites such as drug delivery devices comprising active pharmaceutical ingredients with excipients.[67–73] Moreover, many mechanochemical processes have been successfully scaled-up,[49,56,58,74–80] offering a direct route to translate mechanochemistry from laboratory curiosity to industrial applications.

Indeed, this enormous range of applications for mechano-chemical preparations demands that the processes which govern their transformation must be equally diverse, as shown in the hierarchical diagram in Figure 1. There is some elegance to this complexity: many of the processes which govern mechano-chemical reactions of complex systems can be largely deconstructed into some combination of the elementary processes which occur in simpler systems. For example, mechanical treatment of a single powder particle will still involve geometric distortion of its molecular substituents,[81] and there remains the potential for molecular or atomic electronic excitation/emission processes to occur. This behaviour is clear for example in high pressure experiments of molecular solids, wherein mechanical action of the bulk lattice yields geometric[82] and electronic distortions[83] or excitations[84] at the molecular or atomic level.[85–87] The dynamical stressing (compression or shearing) of solids can also cause chemical species within the solid state to approach each other at high velocities, akin to molecular collisions in fluids. Such dynamical interactions can occur either within a single particle,[88–90] or at inter-particle contacts.[91] These solid state 'molecular collisions' have been suggested as the origin of slip-induced 'hot spots',[88,92,93] or even covalent bond formation.[94]

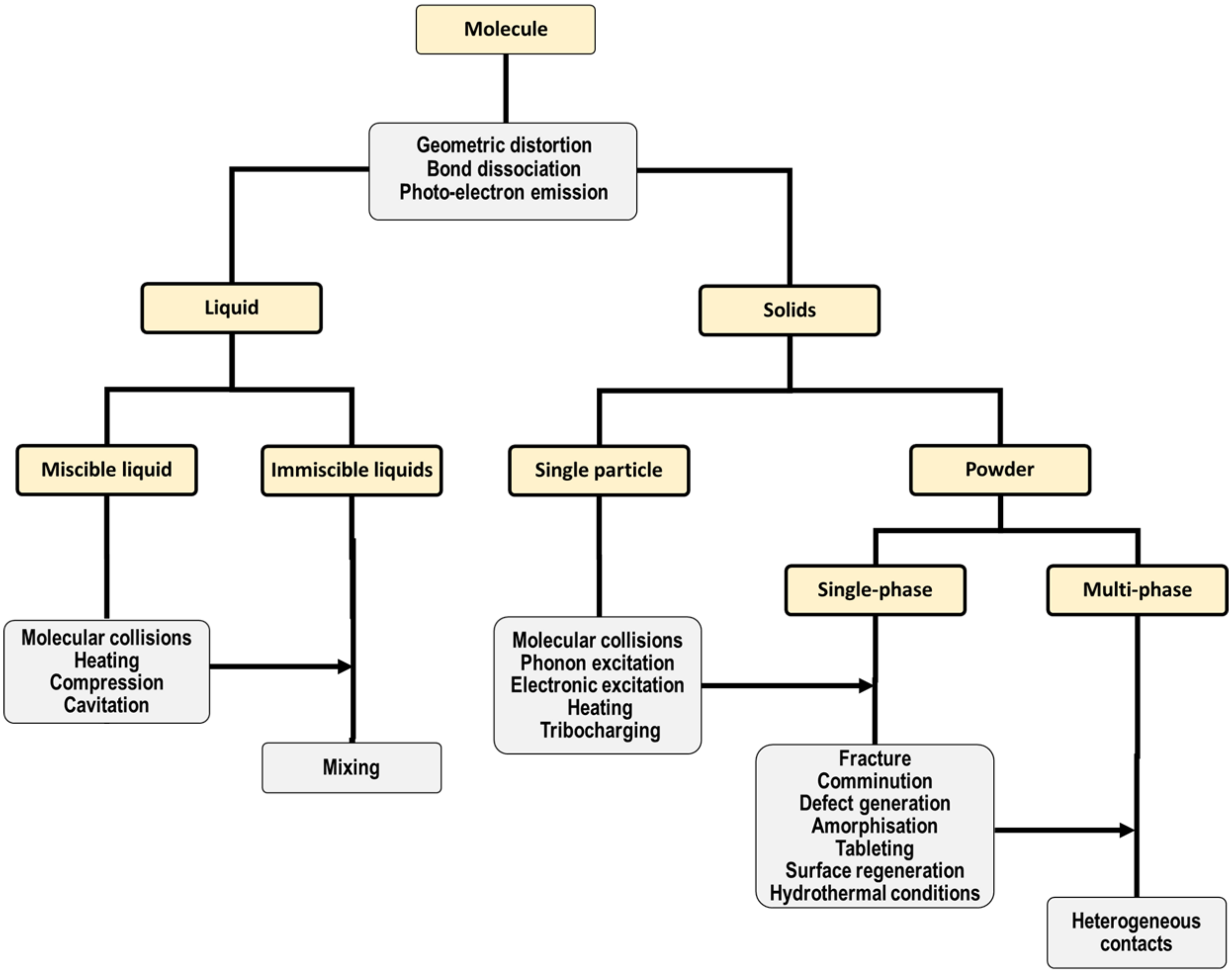

**Figure 1** | Hierarchical representation of the major effects (grey) of mechanical action on different systems, ranging from single molecules to multi-phase solid powder mixtures (gold).

That said, with increasing complexity of the system, many more and new pathways exist by which mechanical action can exert influence. This hierarchical phenomenology is again exemplified by the study of high-pressure phenomena in molecular solids. Mechanical force not only affects molecular geometry but can also influence the intermolecular non-covalent interactions, leading to changes in crystal packing (polymorphism). By manipulating crystal packing, mechanical force thus offers a route to modifying bulk physical properties such as lattice stability, melting temperatures, and compressibility. Thus, although many of the elementary stages of a solution-based chemical reaction may still apply to mechano-chemical reactions, many additional elementary stages must be also considered to fully account for the mechanism of mechanochemical transformations. Identifying, understanding, predicting, and ultimately controlling these pathways leading to the mechanical manipulation of matter is crucial, should mechano-chemical approaches ever become equally controllable as the well-developed aspects of solution and gas-phase reactivity.

Deconvolution of these complex phenomena requires the use of a common language which allows the effective communication of the process being discussed. Only in this way can we hope for a concerted and coherent effort towards elucidating mechano-chemical reaction mechanisms and driving forces, and therefore gain control over these reactions to make them possible to reproduce and scale. Despite decades of mechanochemical research, the need to agree on using certain terms, on how to define accurately and unambiguously the experimental procedures, and how to present the results have been not widely seen as necessary until very recently. For almost a century, the mechanochemistry community remained relatively small, although it covered a diverse range of fields. Researchers knew not only the scientific research of the others but often knew each other personally. The basics of mechanochemistry, as well as the experimental and computational protocols were discussed regularly in original publications, and at the many seminars and conferences. The developments in the field were regularly summarized in thoroughly detailed monographs and reviews, that have now become seminal.[6,20,25,95–103] The foundation of the International Mechanochemical Association (IMA)[104] under the guide of IUPAC in 1989, was an important event that marked the formation of a mature scientific community with a common language and a clear vision of the scientific field.

As an increasing number of groups have begun in recent years to enter the field of mechanochemistry independently of IMA, the quickly growing community has since become scientifically heterogeneous. In contrast to its original composition, the mechanochemistry community is now becoming enriched with researchers from very different backgrounds and expertise, many of them being originally experts not in the solid-state, but in solution-based chemistry. This diversification in its membership has brought with it many new and exciting research challenges, leading to a much greater impact of mechanochemistry than ever before. Yet one cannot ignore the fact that with diversification of the community comes an expanding breadth of specialised scientific languages. As a consequence, the heterogenous community may not always understand each other effectively, or may become increasingly unaware of the mechanochemical knowledge that has been accumulated in early publications, that itself can be perceived as being written in “another scientific language”. There is the real danger that the lack of use of a common scientific language can prevent the community from meeting the challenge of constructing the “Tower of Mechanochemistry”, as it did millennia ago in relation to the Tower of Babel.

## 2. WHAT IS IN A NAME?

Anthropologists argue that humankind evolved due to our capacity to conceive abstract phenomena and communicate these phenomena through complexity of language. In this light, it is no wonder that philosophers have attached such significance to the selection and connotation of words. In Plato’s famous dialogue Cratylus, he argues: ‘a name is an instrument of teaching and of distinguishing natures, as the shuttle is of distinguishing the threads of the web’.[105] The chemical sciences have followed true to Plato’s logic. Chemical reactions are denoted according to the type of energy used to initiate a chemical reaction, their nature is revealed through their name: thermo-chemistry, electro-chemistry, magneto-chemistry, photo-chemistry, and radiation-chemistry. Thus, adequately naming a chemical reaction *requires an elementary understanding of the underlying chemical and physical processes*.

In the early 20$^{th}$ century, Ostwald[106] noted reports that existing nomenclature in the chemical sciences did not fully represent the true nature of all observed chemical reactions, for Carey-Lea demonstrated a unique outcome of thermal- and mechano-chemical reactions in metal halides.[7,8] Correspondingly, Ostwald introduced in his 1919 textbook the term *mechanochemistry* to describe reactions *in any state of aggregation* which are initiated by mechanical force (impact and friction). A more specific term - *tribochemistry* - was subsequently proposed to denote only those chemical and physico-chemical changes which occur *in solids* in response to mechanical energy.[107] The term *tribochemistry* was used widely throughout the 20th century in relation to processes that occur on grinding, ball-milling, comminution, friction, wear, rubbing, and lubrication of solids. With growing diversity of tribochemical reactions, daughter terms became commonplace to facilitate more accurately the communication of the scientific work. These words included tribocatalysis, triboelectrochemistry, tribosorption, tribodiffusion, tribocorrosion, tribotechnology, tribomechanics, tribogalvanics, and tribometallurgy, each introducing specific subfields of tribology, the science uniting tribochemistry and tribophysics.[6,108–110] A primary aims for introducing these "tribo" terms in addition to Ostwald's term "mechanochemistry", was to separate solid-state mechanochemistry from mechanically induced processes that occur in single molecules or liquids. In this way, physical phenomena induced by mechanical action in solids or at their surfaces, including phenomena like mechanical mixing and comminution, were to be denoted with a "tribo-" prefix. All other mechanically induced phenomena were instead to be defined by a prefix "mechano-".

Despite growing popularity of tribological nomenclature, analogous terms prefixed by "mechano" remained in the literature as synonyms also for processes involving solids, including: mechanochemistry, mechanocatalysis, mechanocorrosion, and mechanotechnology. Moreover, the term *mechanical alloying* was introduced to define the process of forming intermetallic compounds and alloys by mechanical treatment of solid components, many of which could not be accessed by any other way than mechanical treatment.[101,111–113] *Mechanical alloying* processes are of great practical importance, which may account for the extensive publications of mechanical alloying studies and, as a consequence, of the very wide usage of this term in the scientific literature. The nomenclature regarding the mechanical manipulation of single molecules has remained much simpler. Only terms prefixed by 'mechano' have been commonly used when discussing transformations of single-molecules induced by mechanical stretching of bonds, e.g. using an AFM cantilever,[52–54,114–116] or when investigating biochemical and biophysical processes.[117–119] Figure 2 gives an idea of the relative frequencies of how the usage of different terms has changed with time. *Tribology* - the term introduced by Jost[120] in the 1960s - remains by far the most popular term to date, while *mechanochemistry* is much less used than *tribology*, or *mechanical alloying*. While the organic chemistry community appears to favour the term *mechanochemistry*, the terms *tribology* and *tribochemistry* are more popular amongst chemical engineers and materials science community. In addition, the term *mechanical activation* is widely used in relation to thermal reactions that are facilitated by mechanically pre-treating.[20,22,121–124]

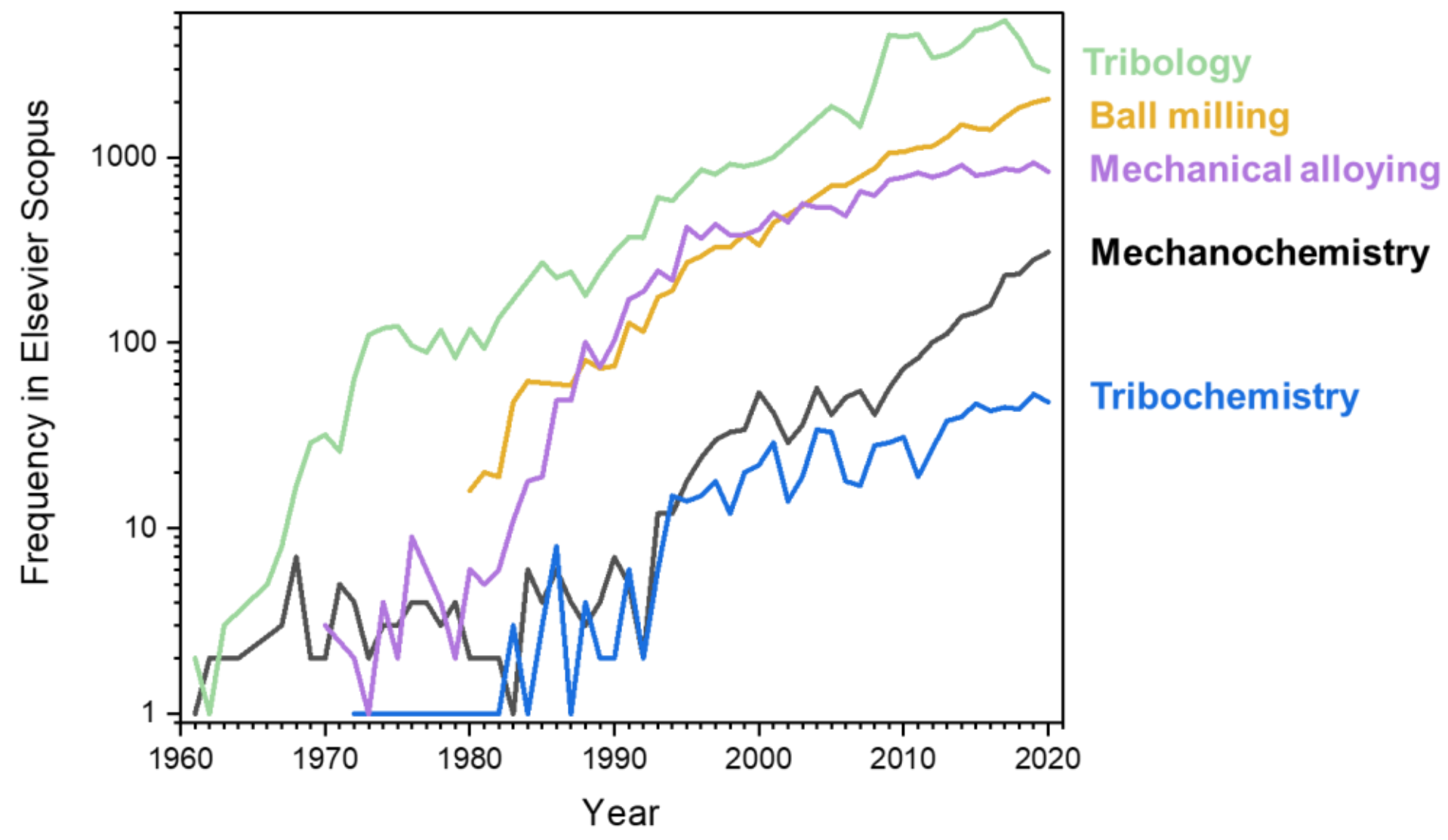

**Figure 2**| The number of papers in Scopus each year in which the terms related to chemical reactions of solids induced by mechanical treatment are used in the title, abstract or as key words.

*What is in a name? That which we call a rose by any other name would smell as sweet.* This famous phrase by W. Shakespeare serves an important lesson for science: the physical world does not depend on our description of it. Of course, our choice of nomenclature has no influence on the physical reality of chemical reactions. This nomenclature does, however, influence our understanding, communication, and formulation of scientific principles. Nature is indifferent to our terminology. Scientists, in contrast, are prisoners to nomenclature. Without consistent and precise definition of scientific concepts, *'the rose'* may not smell as sweet.

Discussions of nomenclature in science has a long history. Most famous, perhaps, are the classifications of species proposed by Charles Darwin. The taxonomic classification of life demonstrates Plato's reflection of a name in understanding the properties and connections between entities. Similar ontological classifications have been popular in chemistry throughout its history. For example, chemists routinely classify *interatomic interactions* according to an abstract definition of *bond order*: *single*, *double*, *triple*, etc. This precise nomenclature allows scientists to directly and unambiguously describe a characteristic of a *molecule* directly by the type of *bond*. Recently, a standard set of bond descriptions were suggested to define particular types of molecular interactions.[125–127] Standardising nomenclature has long been the focus of the IUPAC. More broadly, ensuring that well-defined and well-classified ontologies exist throughout the sciences is becoming increasingly recognized as the route to ensure Findable, Accessible, Interoperable, and Reusable (FAIR) scientific data.[128,129]

So, what is in a name? Would that which we call 'mechano-chemistry' by any other name behave the same? Of course, the physical principles which govern mechano-chemical reactions will behave independent of our chosen nomenclature, but will the conceptual constructions we use to rationalise, discuss, and direct scientific research be so resilient? The idea of complementing the term mechanochemistry by the term *tribology* (tribochemistry + tribophysics), to focus more on the transformations involving solids,[120] is clear and justified. In practice, the more general term *mechanochemistry* appears to survive. Moreover, it is increasingly used as a complete synonym of *tribology*, also when describing ball-milling, grinding, and friction.[130]. This hazy nomenclature would not create much problem if it were always straightforward to identify which mechanically induced *physical* processes in solids were responsible for the *chemical* processes of bond cleavage and formation. This, however, is not the case. Defining accurately by its name the nature of a mechanically induced transformation has serious implications for the type and importance of physical processes which must be considered when seeking to understand mechano-chemical transformations. Moreover, it is critically important to consider that molecules in the solid state cannot immediately react with one-another. Instead, some solid-state phenomenon must first occur which allows collisions not at the level of *particles* but at the level of *molecules.* The nature of this preliminary phenomenon depends on whether the reaction is mechano- or tribo-chemical. For example, whereas mixing and comminution may be dominant preliminary phenomena in mechano-chemical reactions, electrostatic charging or generation of defects certainly dominate many tribochemical reactions.[20] Hence, focus on chemical equilibria presented in terms of solution-chemistry, where "one molecule transforms into another molecule", are grossly oversimplified and neglect many of the critical physical phenomena which separates mechano-/tribo-chemistry from solution chemistry.

Why then is the term *mechano*-chemistry gaining more and more popularity? An intrinsic problem with applying the term *tribo*-chemistry in the specific sense as originally proposed – as opposed to the previously existing term *mechano*-chemistry – is that we must know the mechanism of the process. Specifically, the term tribochemistry should only be used if it is truly a solid-state reaction. This is not obvious, especially for organic compounds, even if we *start* with solids.[131,132] In many cases, the transformation itself, including chemical synthesis, likely occurs in a fluid phase; the possible origin of this fluid phase can be diverse (Figure 3). This is quite often the case for ball milling, grinding in a mortar, or processing in an extruder a mixture of solid organic compounds.[133] Generally, the origin of this fluid phase can be classified as being intrinsic or extrinsic to the reacting system itself. Intrinsic origins include melting (or contact melting)[43,134–141] *s*ublimation of solids,[142,143] or dehydration/desolvation[144] which result from the excess heating of mechanical impacts or bulk heating during mechanical treatment. Where a solids' glass transition temperature is above the milling temperature, one can consider also the formation of transient amorphous phases.[145] It has become very common to explicitly add liquid to a powder mixture to facilitate mechano-chemical transformations, a process dubbed liquid assisted grinding (LAG).[146–148] This is an obvious extrinsic origin of a fluid phase. Even where researchers do not explicitly add liquid, the powder may 'grab' liquid from the environment in a process dubbed inadvertent liquid assisted grinding (IA-LAG).[132]

Although the exact role of fluid phases in reactions that are assumed to occur in the solid-state is not yet fully understood, various roles can be considered. The added fluid phase can certainly influence the mobility of material (by improving rheology, or by completely transferring the process into a solution or a melt).[133,149] Additional to its influence on material mobility, melting can also drive erosion in microparticle impact,[150] or can hinder impact-induced adhesion.[149,151] A fluid, irrespective of its origin, can create hydrothermal conditions,[152] can modify the mechanical properties of the solids via the Ioffe,[153] Roskoe,[154] or Rehbinder[155] effects (i.e. the altering of bulk mechanical properties through surface modification),[156,157] may influence triboelectric phenomena,[158] and can alter the relative stability of product phases through the selective stabilisation of surfaces.[159,160] In many of these cases the process can and should be classified as tribochemical. However, where the process in fact occurs in the fluid phase, it can be no longer classified as tribochemical, nor does it unambiguously qualify to be denoted as a "dry" mechanochemical reaction.

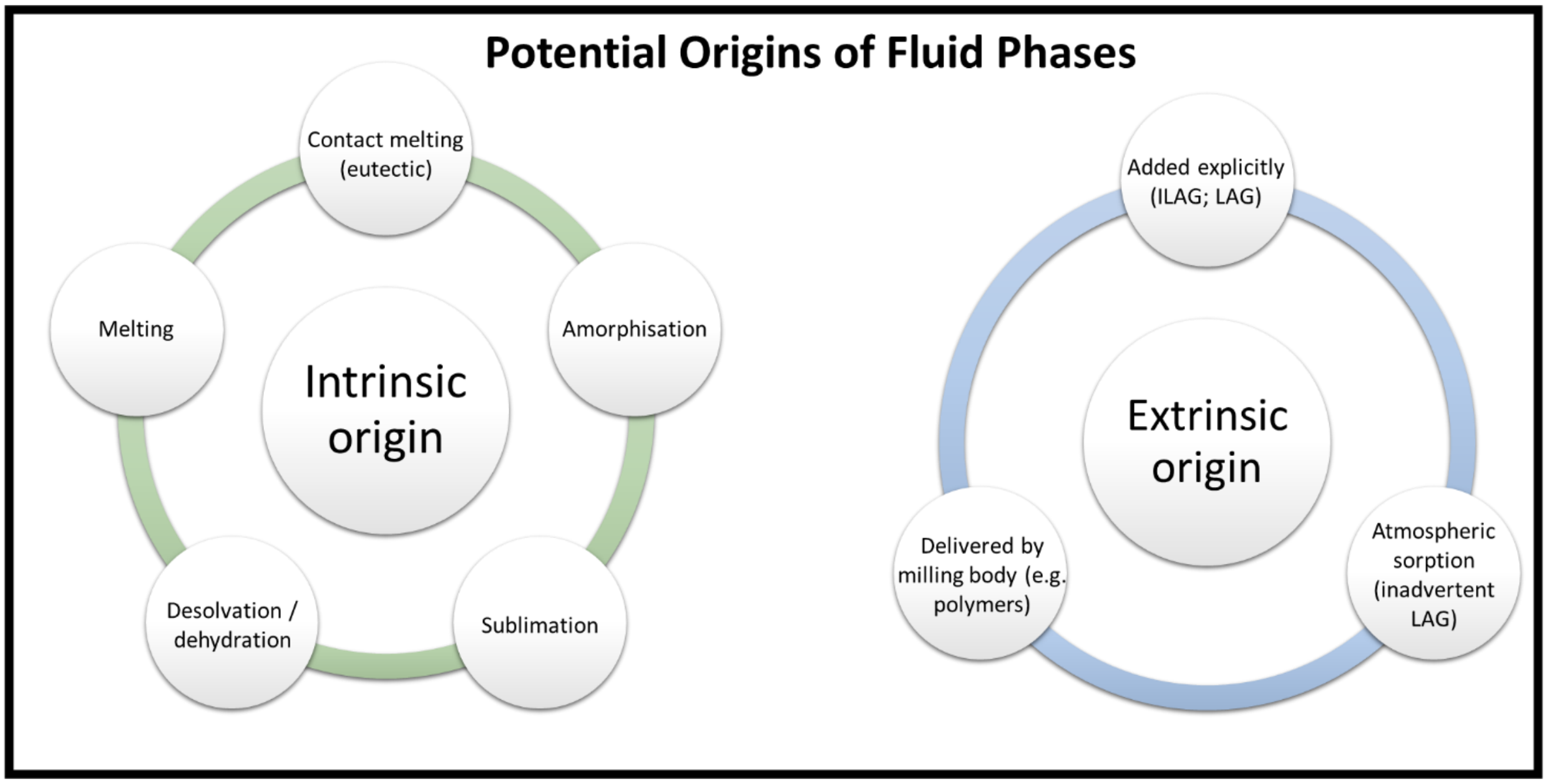

**Figure 3**| The potential origins of fluid phases in mechanochemical systems can be broadly classified as intrinsic and extrinsic, depending on whether they initiate within the solid phases or not.

It is also possible that a reaction in a mechanochemical reactor that starts with solid reactants is in fact not mechano-chemical. This is the case when the chemical or physical transformation itself is not related directly to the absorption of the mechanical energy input. For volatile compounds, the solid reactants do not even require physical contact and can remain separated in space.[143] In this particular case, the reaction is, strictly speaking, neither mechano-, nor tribo-chemical. In most reactions, however, no visible transformation is observed unless the compounds are treated in a mechanical apparatus. Yet, in many such cases the reaction is not mechano-chemical, but thermal in nature; the role of the mechanical processing is limited to facilitating the mobility of the solid reactants, bringing them into contact with each other and/or by removing the solid products which are formed at the surface of the powder particles. Such processes are largely responsible for the success of new mechanochemical reactors such as the Resonant Acoustic Mixer (RAM).[77,161,162] One could term such reactions as "mixing-assisted thermal reactions", as opposed to mechano-chemical reactions. This can be taken as an analogue of stirred solution-phase reactions, wherein stirring does not *cause* the reaction, but simply facilitates the thermal reaction by driving mass transport.

These 'mixing assisted thermal reactions' differ from 'classical' thermal reactions in a few critical ways. 'Classical' thermal reactions in solid mixtures (e.g. in high temperature solid-state synthesis) involve the heating of pre-mixed powders, wherein the powder remains largely unperturbed during the reaction. In contrast, reactions of thermal origin that occur during mechanical treatment are accompanied by dynamically changing local compositions (e.g. from mixing) and fluctuations in particle stress regimes. The mixing can be accompanied by the reduction of particle size (comminution), or changes in their agglomeration state. Together, these phenomena of macroscopic motion of particles, comminution or agglomeration, and the potential for strain-induced dissolution of one phase into another[163–168] yields intimate mixing across lengths of scale. One must also keep in mind that many solid + solid reactions are exothermic, since no entropy is gained during the reaction. Correspondingly, if the mechanical treatment is itself accompanied by significant heat evolution, mechanically-initiated self-sustaining thermal processes become possible.[139,169–174] It follows that when mechanical mixing results in a chemical transformation, it is not clear a priori if the mechanical treatment of particles themselves plays a significant role, or if the particles are "merely brought into contact". Regardless, any 'mixing assisted thermal reaction' will be to a large extent governed by macrokinetics, i.e. heat and mass transfer processes.[136,149,175,176]

It is therefore clear that many so-called 'mechanochemical reactions' are not encompassed by the current IUPAC definition, which states that a mechanochemical process is 'a chemical reaction that is induced by the direct absorption of mechanical energy'.[177] Such distinctions, although semantic, play an important role when considering the types of physical phenomena which may play a role in driving the observed reaction. For example, effects of adiabatic compression, or the generation of vibronically excited states are unlikely to play a role in 'mixing-assisted thermal reactions'. On the other hand, slow nucleation and crystal growth – which presumably dominate such thermal reactions – are very different from the fast cooperative interfacial propagation processes which can be expected for 'true' mechano-chemical reactions in which mechanical energy is directly transferred into high-level vibrational or electronic excitations.[178–185]

By considering the possible intrinsic origins of fluid phases it becomes clear that some compounds are more likely to give rise to tribo-chemical transformations. For example, solids with high melting temperatures (primarily inorganic solids) will not melt or sublime during ball milling and therefore will most likely react tribo-chemically. In contrast, materials with low melting temperatures (e.g. most organic or coordination compounds) which are likely to melt during or as a result of mechanical treatment cannot be even strictly classified as mechano-chemical, though tribo-chemical effects (e.g. electrostatic charging) may be still be of importance at the elementary (molecular) level.[11,186] This represents a critical difference between inorganic and organic 'solid state' transformations under mechanochemical action.[133] Important phenomena such as triboelectric charging[158,187] and the mechanical generation of exposed surfaces[188,189] or defects are equally likely to occur in inorganic, organic, and polymeric compounds. It is probable that such phenomena involving defect formation[190] play a central role in most solid state mechanochemical transformations, even if they are often overlooked and subsumed by explanations of "mere mixing". Mechanically generated defects can range from radicals and the isomerization of molecules to extended stacking faults, dislocations, and formation of shear structures. For example, it is known that grinding initially leads to particle size reduction, and ultimately (after a critical grinding or comminution limit is achieved) to the accumulation defects within the solid structure.[191] This change in stress relaxation mechanism can lead to deep mechanical activation thereby greatly affecting the reactivity of the solid. Such effects have been suggested as being responsible for the extended induction periods observed in some mechanochemical reactions.[192]

Selecting a proper term for a process that occurs in response to mechanical action is intrinsically challenging; the correct term can be only given after the mechanism for the reaction has been established. Moreover, there is significant likelihood that many reactions will need to be reclassified as our understanding of the mechanisms of mechanochemical reactions expands. To paraphrase N. Copernicus, we know what we know, but we do not yet know what we do not know. This reality is of course impractical and indeed superfluous for most researchers who are more interested in the outcome of mechanical treatment rather than in the detailed mechanism of the transformation. What *is* important, however, is that the process can be reproduced by others based on the original, recorded description. This requires effective and accurate communication of the mechanochemical protocol used, with meaningful descriptions of all the parameters that are known to influence mechanochemical reactions. Hence, although we cannot know *a priori* how a seemingly 'solid + solid' reaction will occur, we can know for sure how we treat the sample and analyse the outcome. It is also important to determine and record the appearance and state of the sample at the start and end of our treatment. It is this information rather than a name itself that must be reported and controlled as carefully as possible. Only in such a way can we hope to identify under what conditions solids react, and how to implement this technology most effectively.

## 3. The Breadth of Mechanochemistry

If we define a reaction as thermochemical, it is sufficient to indicate the temperature at which it occurs. If the reaction is not isothermal, a protocol of modifying the temperature with time is required. For a photochemical reaction, energy, polarisation, the intensity of light, the spatial characteristics of the irradiation (uniform, local, one-sided, etc.) must be defined. In contrast, where irradiation is discontinuous, the duration and frequency of light pulses must be stated.

For a mechanochemical transformation the type of mechanical action, experimental conditions, the composition and appearance of samples are much broader than for a thermal, or a photochemical reaction (Figure 4). In fact, the very questions as to 'how we treat a sample' and 'in what state the reactant and product phases exist' are not easily defined. Increasingly, new features are being identified to be crucial for determining the reaction pathway of mechano-chemical transformations. For example, the ability for starting reagents to exist in different solid forms (polymorphs, polyamorphs, particles of different size and shape) is unique to the solid state. However, the starting forms of the solid reactants are often not reported in literature, under the assumption that they will be modified by treatment anyway. However, a starting polymorph can play a non-negligible role in determining the outcome of a mechano-chemical transformation.[193] Evidence has suggested also that the particle sizes of reagent materials can influence the mechanochemical transformation.[194] Similarly, the presence of unobservable contaminants (seeds)[195] or crystal defects[196] can be critical to the success or failure of a mechanochemical transformation. It is without doubt that more influential parameters will become known with further studies of mechanochemical reactions. At present we can only make every effort to record and describe as many parameters as possible regarding mechanochemical transformations and remain ready to adapt to new developments as they emerge.

The type of mechanical action used to promote mechano-chemical transformations has a clear influence on its outcome.[197,198] This has become increasingly important owing to the rapidly expanding repertoire of mechanochemical reactors. These ranges from the original mortar and pestle and ball milling (e.g. planetary and vibratory) to twin screw extrusion (TSE), Resonant Acoustic Mixing (RAM), diamond anvil cell (DAC) technologies, and atomic force microscopy (AFM). The choice of mechano-chemical reactor has significant influence on *how* the stress is applied to the sample in the first place Figure 4.

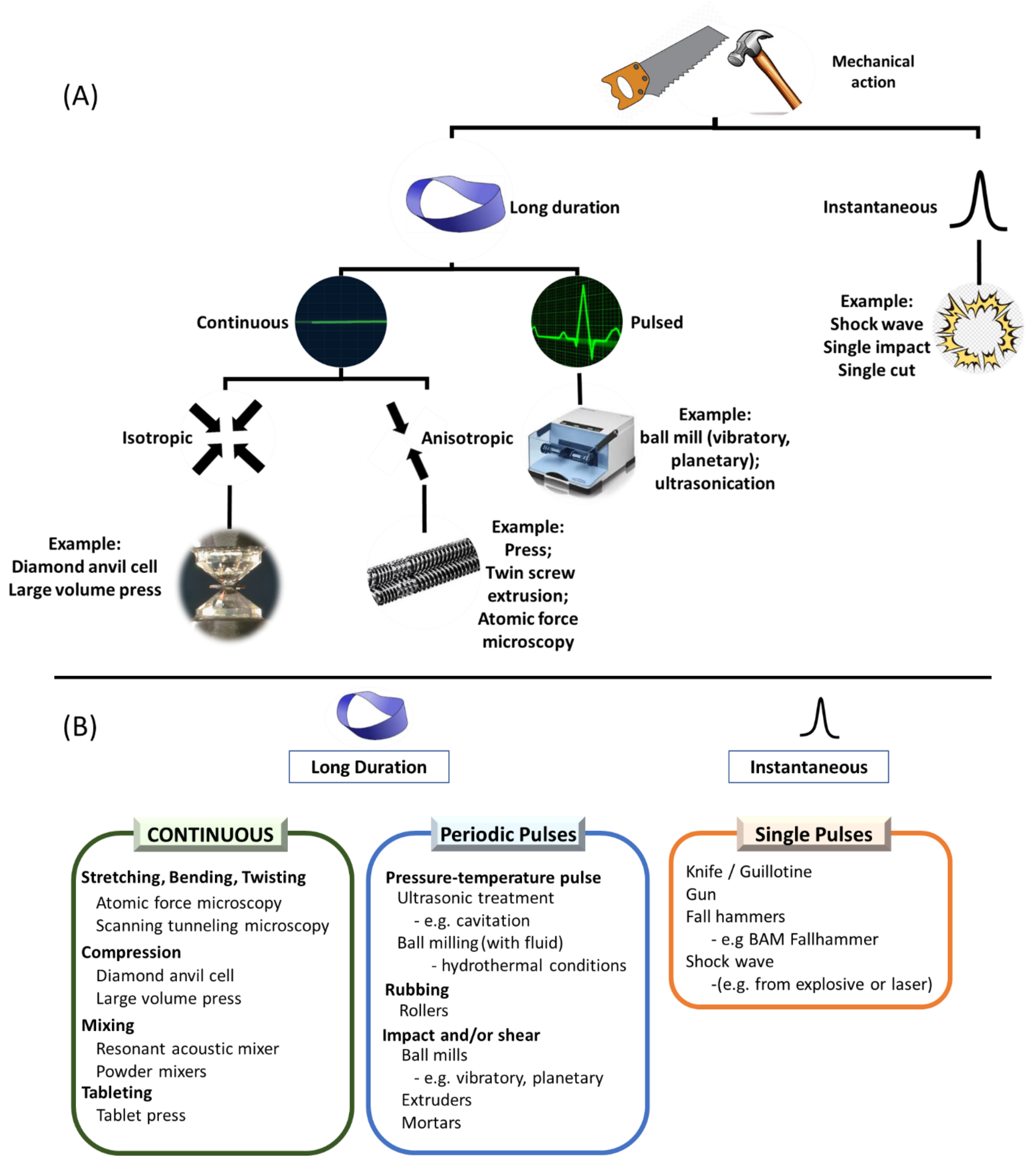

**Figure 4**: Representation of the breadth of mechanical action which are frequently used in mechanochemistry. (A) A broad classification of the types of mechanical action encountered in mechanochemical reactions. (B) Examples of common devices under each category

The temporal evolution of mechanical action can be regarded as a first parameter to consider when distinguishing mechanochemical regimes of treatment. If a shock wave or a single pulse (e.g. a drop hammer, a knife, a gun, or a jet mill) is used, there is a single excitation event, followed by relaxation of the material via various channels (Figure 5A).[20] For example, this is important in the case of the mechanochemistry of explosives,[178] wherein the mechanical impact is suggested to induce super-heating of lattice phonons,[199] which ultimately yields a chemical response *via* dynamical metallization.[185] Chemical transformations at a crack tip fall also into this category. As the crack propagates through the material, the exceptional transient stresses generated at the crack tip induce highly metastable states which, after the crack dissipates, relax through chemical recombination.[200–205] During explosive initiation and detonation, chemical changes may in fact precede heat, such that the process is athermal and thus truly mechanochemical.[206] The rate of crack propagation (akin to the magnitude of a shock wave in explosives) dictates the magnitude of metastability and can even influence the relaxation product, i.e. the chemical products of the mechanochemical decomposition.[207,208]

Typically, treatment of a sample by a single mechanical pulse is used in model studies, when the details of the effect of the mechanical action on the sample are studied - light, electron, radicals, or gas emission, heat evolution, propagation of the deformation wave, generation of phonons, a hot spot formation, etc. For example, the effect of controlled single impacts on the decomposition of Ag oxalate to form Ag nanoparticles has been recently investigated by Delogu and colleagues.[209] Studies on single pulses are particularly well suited for the *in situ* study of fast transformations, and indeed form the basis of safety technologies e.g. in testing explosives (*c.f.* the BAM Fall Hammer or the Rotter Impact Device).

If the mechanical action is long-lived, however, the relaxation event is altered, and the complexity of the problem increases. **Long-lived** action can itself be divided as being **continuous or pulsed**. In the former, the stress, is applied and maintained. The stress must relax on a so-called force-modified potential energy surface (FM-PES), which can be significantly different than the surface of the fully relaxed material. Moreover, the rate at which the stress is applied can affect the effective FM-PES upon which relaxation occurs.[210] Atomic force microscopy (AFM) can be used to stretch continuously a selected bond in a molecule. AFM and other dedicated stretching or bending device can be also used to deform a macroscopic object - a single crystal, a polycrystalline sample, a polymeric fiber - elastically or plastically It can be used to study physical and chemical transformations in such strained samples.[211–214] Alternatively, one can subject a sample to hydrostatic compression (a small amount in a diamond anvil cell, or a larger amount in a large volume press) and study either the transformations induced by compression itself, or the effect of compression on thermal, or photochemical transformations[85,215]

Most devices are based on repeated mechanical treatment of samples in the form of pulses. In the case of **repetitive dynamic stressing**, relaxation of the stress occurs on the unstressed PES. However, depending on the intensity, pulse shape, duration and frequency of pulses, the relaxation may not be complete. This offers a potential means to accumulate energy with successive pulses.[16,192] Additionally, the relaxation channel (Figure 5A) can change during the course of the reaction, for example as the result of reduction in particle size towards the grinding limit.[191] The rate of pulsing can vary significantly, from *e.g.* 30 Hz in a conventional ball mill to tens of kHz using ultrasound radiation.[48] The types of mechanical action can be also different, including impact, shear stress, friction, rubbing, and cleavage, or any combination thereof, Figure 5B. Moreover, different types of action can be applied to fractions of the sample located at different sites within the same milling jar,[198] or can vary between successive pulses. The type and variability of mechanical action does not depend only on the choice of the mechanoreactor, but also on (1) the protocol of operation for the mechanochemical reactor,[100] (2) the nature of the sample (e.g. its rheology),[194,216] (3) the presence of additives (e.g. liquids, polymers), and (4) the nature the milling bodies[217] (e.g. their material, mass, density and size and the total number). Each new pulse can interact with the same particle in potentially different geometry, or with another particle altogether. Under such dynamic stressing conditions understanding of, and control over, the type of mechanical action is generally limited. Model devices have been constructed in attempts to minimise the variation between reaction sites and successive impulses, and to monitor step-by-step the evolution of systems under consistent and repetitive stress.[197,218] By using such model devices it has been possible to attempt to rationalise the outcome of more complex mechanochemical reactors, such as vibratory ball mills.[140,198] It is likely that at different stages (comminution, mixing, generating defects, reaction itself) effective mechanochemical transformations require that a sample be processed in different machines which are optimized for these particular phenomena. Developing model devices to simulate and study systematically mechanochemical transformations is exceptionally important rationalize how and why mechanical treatment exerts its effects. To this end, new model devices capable of probing at the nanoscopic level the evolution of solids exposed to mechanical treatment are being also envisioned.[219]

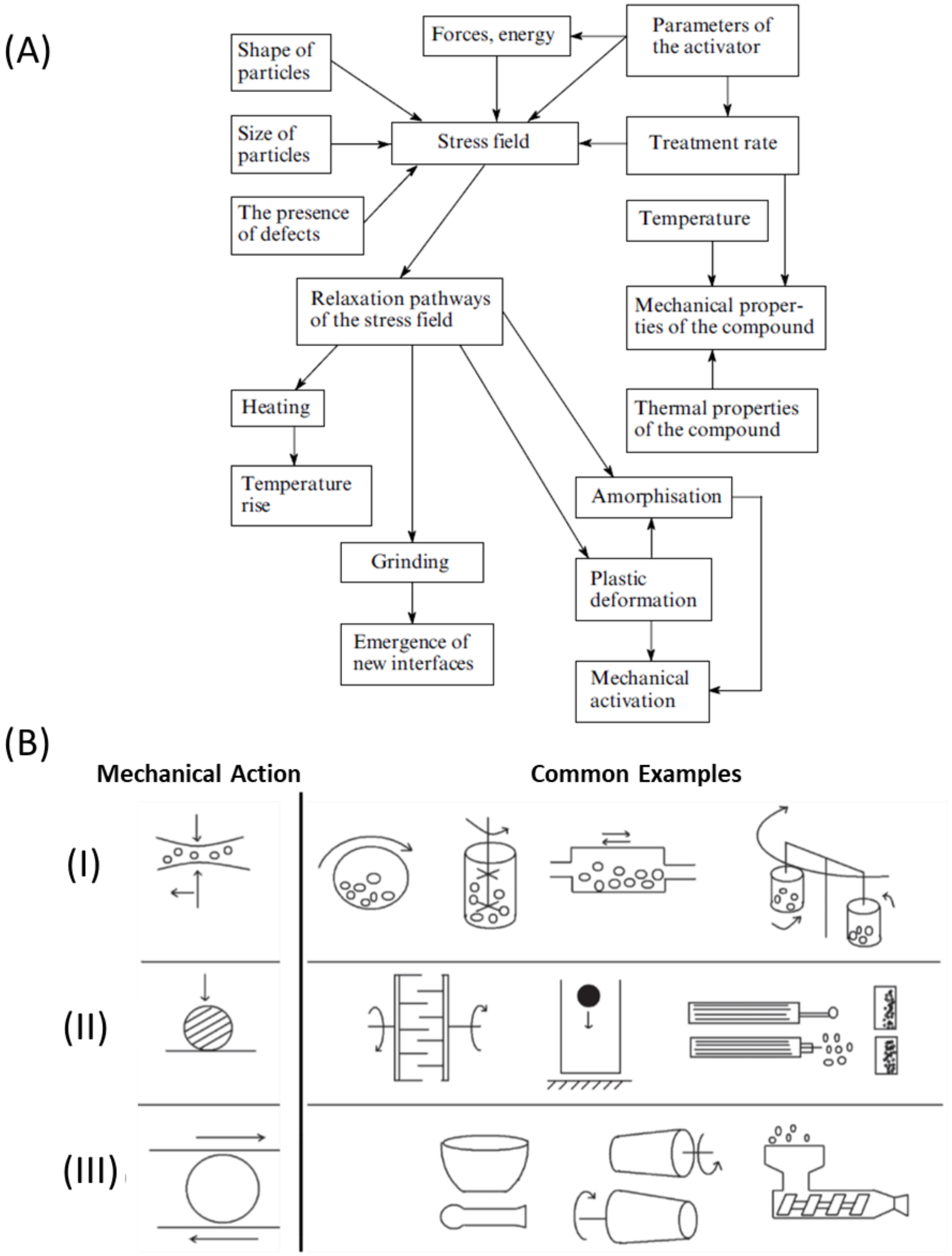

**Figure 5|** Different influence of mechanical action on solids. (A) Flow diagram for the formation and subsequent relaxation of stress which are involved in mechano-chemical reactions in solids, adapted from Ref [20] (B) Schematic representation of the breadth of mechanical action experienced in different types of mechanochemical reactors. (I) combination of impact with shear/friction in various types of ball mills and attritors (from left to right) rotational mill, attritor-stirring ball mill, vibration mill and planetary mill). (II) Dominated by impact in (from left to right): pin mill, fall hammer, jet mill. (III) Dominated by shear/friction (from left to right) mortar and pestle, rolling mill, and extruder. Figure adapted from Ref [149]

Identifying the type of mechanical regime, and indeed the type of mechanical action is not obvious but in exceptional circumstances. In the general case, no sharp boundaries exist between the regimes of mechanical treatment. For example, in ball milling experiments the powder is constantly relocating through the jar. Consequently, the frequency and intensity of pulsed loadings is not constant. Indeed, an apparently 'repetitive dynamic' regime may be in fact more accurately described by 'single pulse' regime of reactivity. In contrast, if powder is tableted during impact, material remains under some compression. Moreover, when tableted, the type of mechanical impact (e.g. from free impacts to restricted impacts) or potential for shear (e.g. particles against the jar walls) are greatly affected. In this case, the regime of mechanical treatment changes unexpectedly. The stochastic and time-dependent formation of 'reactive' and 'unreactive' contacts, which are affected by mixing, can be also unpredictable. For example, in a two-phase reaction, only mechanical treatment of contacts which involve both phases (i.e. a heterogeneous contact) will react.

In contrast, homogeneous contacts will not undergo reaction. Thus, the relaxation of mechanical stress depends also on the structure of the inter-particle contacts, which vary unpredictably with time. Even if one start with treating a single phase (for example, when a polymorphic transition, an amorphization, or a decomposition are studied), as soon as the product starts being formed, the system becomes at least "binary", or even contains more different phases. Therefore, the next impacts can hit not the yet unreacted particles, but the already formed product. This has dramatic consequences on the kinetics of the transformation.[220] Even such "single phase" mechanochemical transformations turn out to be strongly controlled by macrokinetics, i.e., mass and heat transfer, comminution, mixing and aggregation.[176]

## 4. A Need for More?

A true designation of a mechano-chemical (or tribo-chemical) reaction requires a thorough understanding of its mechanism. In all but a few cases, this understanding is far from being achieved (See Section 3). We must therefore ask the question: Is an understanding of the mechanism a prerequisite to garnering control over potentially mechano-chemical reactions at all? We might not need to know what *exactly* is occurring, but we *must* understand collectively what actions are being done, and how to describe these actions amongst the community. To pool knowledge, and ultimately achieve the mechanistic understandings required to develop robust and meaningful *names*, it is vital that the strategies being used across the domains of mechanochemical reactions be effectively communicated. This approach to developing a common language in reporting science has been crucial for example in progressing the field of crystallography. The creation of a *Crystallographic Information File* (CIF)[221] facilitated the deposition and automatic processing, visualization and validation of crystallographic structural data. Additionally, the CIF format facilitated the wide adoption of common standards for good practices in the collection and processing of structural data by highlighting to researchers which experimental parameters must be controlled and reported, such that anyone can validate and reproduce the results.[222–224] During its inception, researchers understood that the field remained in development, both in terms of technology and in understanding which parameters were required for the effective communication of crystallographic data. Hence, the CIF file was established to be dynamic and 'naturally evolving'. The process continues still, including increasingly difficult cases such as structural data from powder diffraction, for modulated structures, and only very recently expanding to include high-pressure data.[225] This situation is very similar to that which faces mechanochemistry today.

It is certain that no universal set of parameters or nomenclature can span the entirety of mechanochemical research. The diversity in mechanical action, stressing regimes, phases present in the system as reactants, additives, reactor materials, or inadvertently penetrating water or impurities, to list just a few parameters that need to be controlled, are too diverse. Yet, this is again synonymous with crystallography, wherein different diffractometer set-ups, collection strategies, sample types, all require a specific set of parameters to be recorded. It is instead likely that effective communication of mechanochemical sciences must evolve to include as many standard parameters (e.g. additives and reactant composition) as possible, whilst allowing the flexibility to dictate mechano-reactor specific information.

A significant step in this direction to standardize nomenclature is credited to Hanusa *et al.*, Figure 6.[206] The addition of energy into chemical reaction schemes is easily denoted for traditional methods. For example, the photochemical activation is denoted by $\hbar\omega$, describing the addition of a quantum of light. Similarly, heating is denoted by Δ. The "three-balls" symbol was suggested to be used to denote any mechanochemical transformation [226], and the simplicity of this symbol has thus far rendered it popular amongst a significant part of the organic mechanochemistry community.

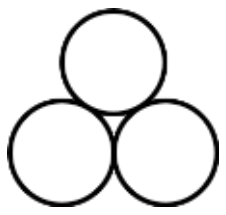

**Figure 6**: Symbol proposed for mechanochemical activation by Hanusa *et al*.[226]

This symbol represents an excellent step towards realizing the need to discern clearly transformations resulting from mechanical treatment from those initiated by thermal, photo, or electrochemical stimuli. However, despite its popularity, the 'three ball' symbol appears to have drawbacks when it is supposed to be used as a universal denotation for mechanically induced reactions.

Culturally, the symbol coincides with that of an ancient religious symbol that has existed since prehistory and is found throughout the world. A nearly identical symbol was subsequently adopted as the symbol for the "Banner of Peace", symbolizing the protection of cultural property in times of war.[227]

As was illustrated in the previous Sections, mechanochemistry differs from other branches of chemical science in that there exist many ways in which the mechanical action can be applied to the system. These different mechanical actions frequently yield different results. Material treated in a vibratory ball mill will be subjected to very different mechanical forces than the same powder treated in a planetary ball mill. Treatment in an extruder, jet mill, or in the rolling mill will be even more different. A single symbol cannot be therefore sufficient to define a general regime of mechanical treatment. Moreover, the presentation of milling balls is not relevant for example in many organic syntheses wherein excellent and scalable results were obtained not by impact in ball mills, but by shear in mortars, extruders, or rollers.[60,75,78,228–230] In fact, a vast majority of efficient and scalable mechanochemical types of treatment are "ball free" (ultrasonic, RAM, TSE, jet milling, pin milling, rolling-milling, grinding in a mortar, to name just a few).

### 4.1 The Need to Specify Mechanical Action

The specific reactivity of materials under different types of mechanical action has been known for decades. Most famous, perhaps, is the reactivity of energetic materials, whose friction and impact sensitivities can differ enormously. Correspondingly, specialists in energetic materials research make explicit effort to ensure that the type of mechanical action is reported alongside the magnitude of perturbing energy; this reporting and testing procedure has been internationally standardised.[231] It is not surprising to find that the behavior of other organic systems is also mechanoreactor-dependent. For example, in the case of the "piroxicam + succinic acid" system, a co-crystal is formed from the two components on repetitive impacts by a falling ball.[197] On the contrary, piroxicam : succinic acid co-crystal decomposes into the component on grinding when shear dominates.[197] Uninterrupted ball milling of a mixture of glycine and malonic acid gives a different product as compared with the outcome of a treatment that is interrupted intermittently and where the sample is mixed manually.[140,198] The work by Belenguer and colleagues demonstrates clearly how the selective choice of ball milling protocol can be used to target with great reproducibility the polymorphic form of organic solids.[159,160] The epsilon-polymorph of chlorpropamide can be transformed into the alfa-polymorph only on cryogrinding at the liquid nitrogen temperature, whereas no transformation is observed on ball-milling at room temperature.[232] Mechanical treatment of the beta-, gamma-, delta- polymorphs of chlorpropamide gives different products, depending on the choice of the starting form, the type of treatment (grinding in a mortar, simulated impact and shear, or during ball milling) and the presence of a small liquid additive.[193] Liquid assisted Resonant Acoustic Mixing of carbamazepine and nicotinamide gives different products depending on the magnitude of acceleration.[161] Such examples are not limited to organic reactions. For example, depending on the rate of crack propagation in crystalline nitrates and chlorates, different products are formed on their mechanolysis (i.e. mechano-chemical bond rupture and formation of radicals).[207] Similarly, the dissolution rate of quartz in hydrofluoric acid was also shown to depend critically on the type of mechanical treatment, with vibratory milling yielding slower dissolution kinetics than jet milling.[6] Similar effects are observed for the formation of a Zn – fumarate metal-organic framework, which forms the tetrahydrate upon intense treatment in a SPEX-8000 mill, but a pentahydrate upon restricted impact treatment or laboratory vibratory ball milling.[233,234]

### 4.2 A picture is worth a thousand words: pictographic representations of mechanochemistry

It is thus increasingly clear that many often-over-looked parameters can influence on a mechano-chemical transformation. Additional to the choice of mechanoreactor,[197] this includes the initial polymorphic form,[193] the inclusion of additives (solid, liquid, or gaseous),[160,195,235] the bulk temperature of the mechanoreactor,[236] the atmosphere in which the reaction occurs,[132] the rate of mechanical stressing,[161] the mass and material of the milling balls and jars.[217,237,238] Certainly, many additional and unsuspecting parameters can also exert significant influence over control in mechano-chemistry.

Despite this, modern mechano-chemical literature typically describes the applied mechanical action in general terms. For example, transformations are often described as occurring under 'ball milling conditions'; in many recent papers the mechanochemical reactions are denoted with a general symbol as in Figure 6. The onus thus remains on the careful reader to identify the type of mechanoreactor, and subsequently the main type of mechanical action being applied. This information can be given in a concise form in the Experimental, or, even more difficult to see immediately, in Supplementary Material. A considerable amount of crucial information is therefore not immediately visible, even if available in the publication. In the absence of a "check list" indicating all the important experimental details that *must* be reported, it is also not uncommon that crucial information is missing. This is often the case, when mechanochemical work is being reported and published in general interest chemistry, physical chemistry, synthetic organic chemistry journals, and not so in the specialized solid-state, materials science, chemical engineering, or tribochemistry journals.

We propose a demonstrative model for a more elaborate, albeit concise, description of a mechanochemical reaction. It is our hope that such a pictographic "nomenclature" will facilitate clarity of communication whilst encouraging researchers to share vital information regarding mechanochemical syntheses. In the model nomenclature, emphasis is on the type of mechanical action, which is placed at the centre of the symbol. Additional parameters which are currently known to affect the outcome of mechanochemical transformations are indicated systematically in the surrounding space, Figure 7. Such a formalism could make great strides towards ensuring comparable, consistent, and reproducible reporting of mechanochemical reactions.

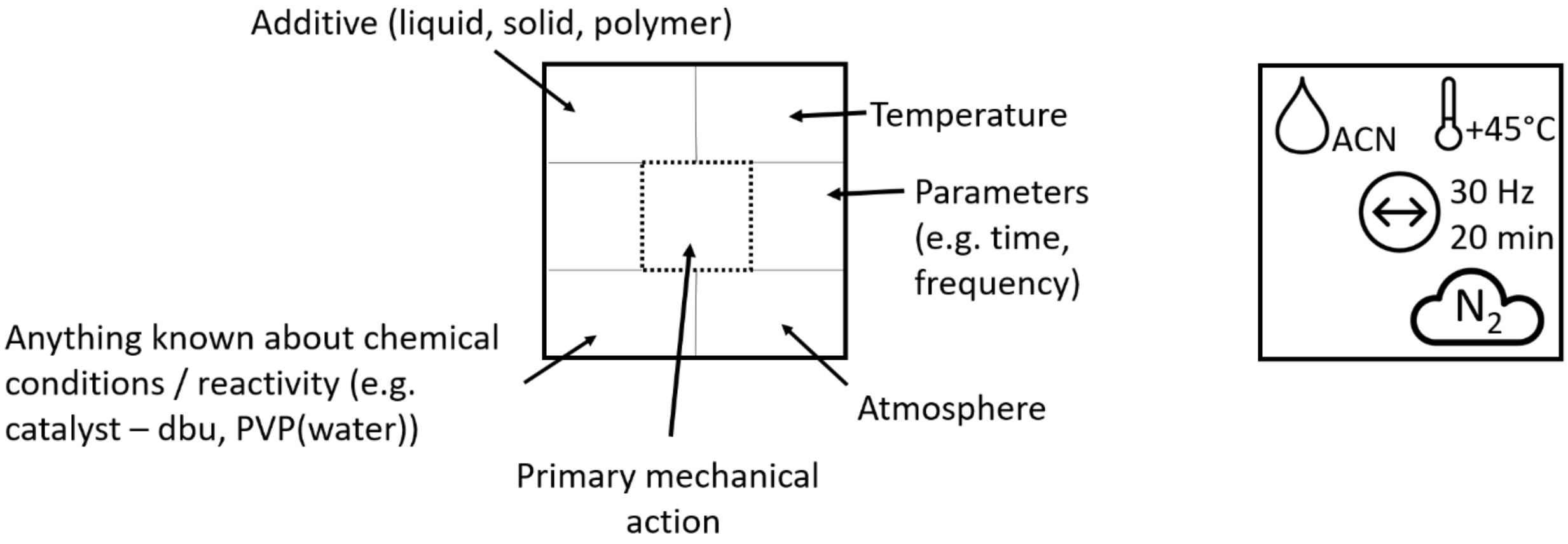

**Figure 7|** Mechanochemistry at a glance. Left: Schematic representation of the structure of the proposed mechanochemical reaction symbol. Types of symbols see Table 1. Right: Example of the symbolic representation of a mechanochemical reaction performed in a vibratory ball mill for 20 min at 30 Hz under LAG conditions using ACN at 45°C under N2 atmosphere.

In this representation, parameters are given that are known to affect the results of a reaction. These include:

*Additive:* a solid, liquid, or polymer that is added to the mixture, in order to facilitate the physical transformation, and may or may not be chemically involved in the transformation. This includes processes such as liquid assisted grinding (LAG), ionic liquid assisted grinding (ILAG), polymer assisted grinding (POLAG), or the addition of polymer to affect rheology. We suggest that the symbol contains a short acronym to label the additive, for example acetonitrile – ACN.

*Temperature*: In cases where the vessel temperature is explicitly controlled (either heating or cooling), the temperature can be indicated here. We suggest that the temperature is denoted alongside the associated symbol.

*Known chemical reactant*: In addition to denoting the additive, any species which are known to be chemically involved in the reaction should be denoted. This includes catalysts, including cases where the milling bodies act as catalysts themselves. We suggest that a short acronym be added to label the reactant species, for example 1,8-Diazabicyclo[5.4.0]undec-7-ene – DBU.

*Atmosphere*: An indication as to the atmosphere within the milling vessel should also be specified. This may be ambient atmosphere (atm), dry air (dry), or any other atmospheric conditions.

*Primary mechanical action*: A symbol should be selected to best represent the type of mechanical action being implemented. Additional information regarding reaction time and mechanoreactor conditions (e.g. frequency) should also be given.

To facilitate the consistent use of this system, we have established a preliminary set of suggested symbols, Table 1. The set of symbols can be easily extended e.g. for new types of mechanoreactors or additives.

**Table 1**: Potentials symbols to be used for each major type of mechanical action.

| Primary Mechanochemical Action | | Additives | Temperature Control | Atmosphere |
|---|---|---|---|---|
| Vibratory Ball Mill | Twin Screw Extrusion | Neutral Liquid | Heating | Ambient Atmosphere |
| Planetary Ball Mill | Sheering | Ionic Liquid | Cooling | Dry Air |
| Manual Grinding | Impact | Solid Additives | | Nitrogen Atmosphere |
| Mixing e.g. RAM | Ageing | Polymer Additives | | Humid Atmosphere |
| Tensile Stretching | | | | |

**Example graphical representations for mechanochemistry.**

Let us consider a few examples to illustrate, how much clearer experimental conditions can be at a first glance if we use these pictographic descriptions.

As a first example, we compare the frequently used mechanochemistry symbol (see Figure 6) to the alternative nomenclature proposed here, with a simple ball mill grinding cocrystal formation, Figure 8.[194] [194] In the original nomenclature, no additional information is visible as to the experimental conditions. In contrast, the proposed new graphical nomenclature immediately allows the reader to identify this as a vibratory ball milling reaction at 30 Hz for 20 min, under atmospheric conditions. The LAG conditions are also visible in the new reaction, in which the water originates from the OAD.[10]

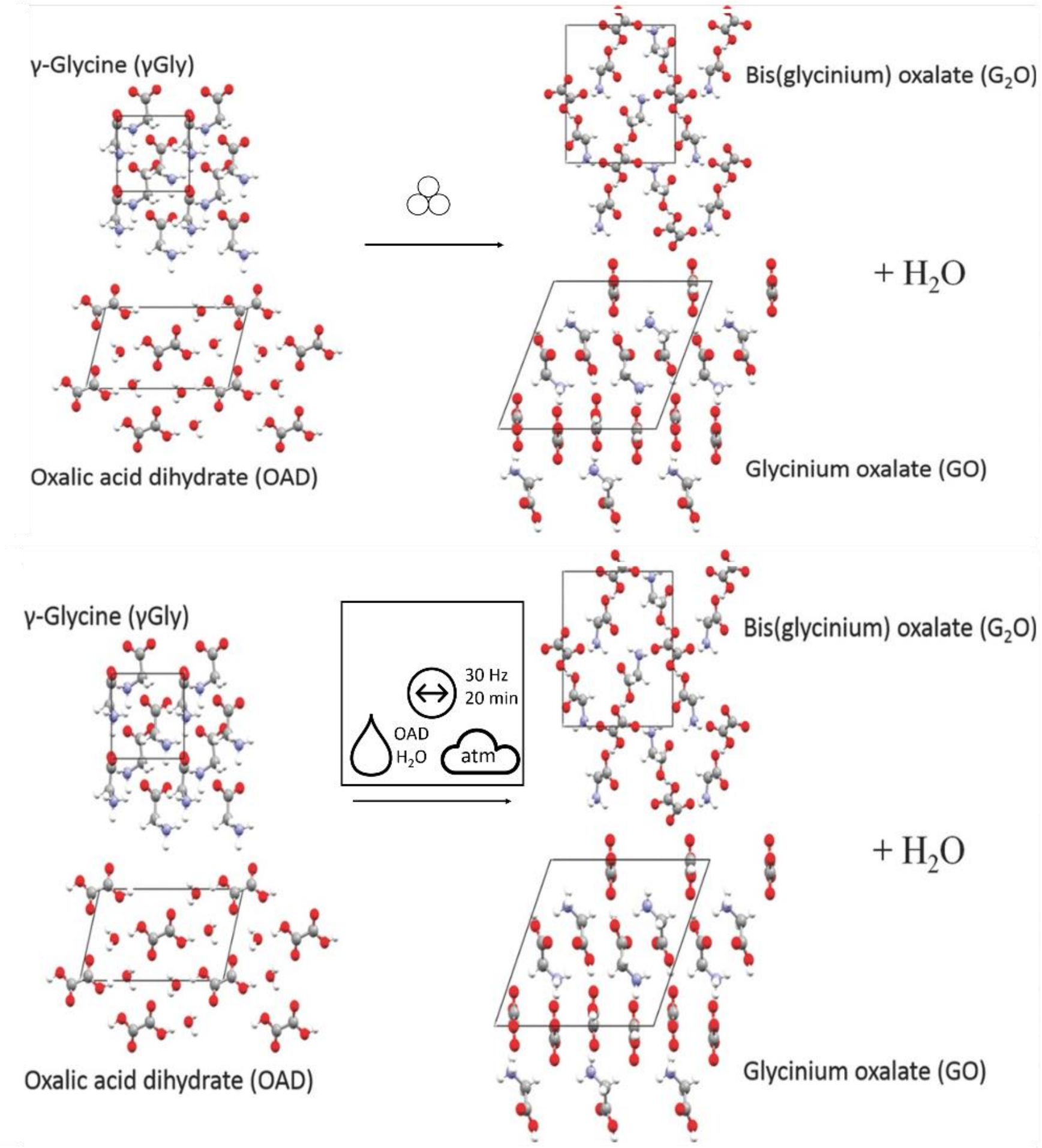

**Figure 8|** Comparison of a simple vibratory ball milling co-crystal formation. (Top) The original nomenclature[194] and (Bottom) the proposed nomenclature.

It is next worth considering a more complex reaction – that of the vibratory ball milling-induced disulfide exchange reaction, Figure 9.[192] In the original publication, the ball milling reaction was denoted as in Figure 9A-B. The catalyst (here dbu) is shown above the reaction arrow, as well as the LAG conditions in ACN. No further information is given as to the nature of the energy source, the physical state of the dbu, the atmospheric conditions or any temperature control. Instead, the reader must find this information for themselves. All of this information is instead visible pictorially using the proposed general nomenclature for mechanochemical reactions. It is immediately evident that the reaction is conducted by vibratory ball milling at 30 Hz under atmospheric conditions, over 30 min for LAG and 45 min for neat grinding (NG), and in the presence of liquid dbu. Furthermore, it is obvious that the two reactions differ only in the presence (or not) of the liquid additive, ACN. Hence, with little extra effort, an enormous amount of additional information (much of which was not present in the original manuscript) is now immediately accessible to the reader.

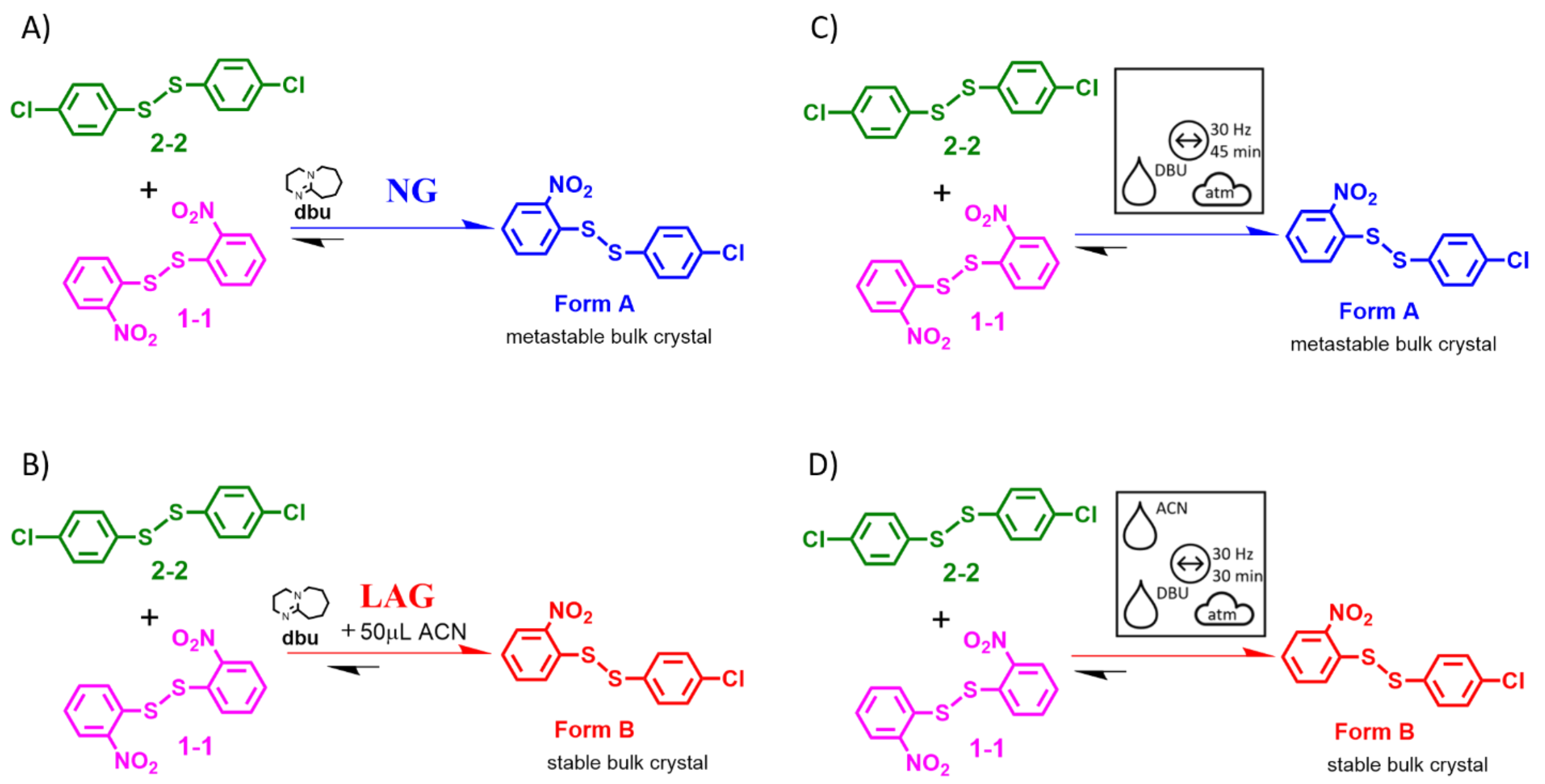

**Figure 9|** Reaction schemes for the mechanochemical disulfide bond formation. (A,B) adapted from literature, Ref[192]. (C,D) The same reactions using the proposed general symbols for mechanochemistry.

An additional feature that is missing from current mechanochemical literature is a schematic approach to denoting the type of additive being used. This is of primary concern when comparing the effect of the common approaches of liquid assisted grinding (LAG), ionic liquid assisted grinding (ILAG) and polymer assisted grinding (POLAG). To demonstrate the strength of the proposed nomenclature, we selected representative examples of each method and show the immediate recognition of these three concepts, schematically, Figure 10 Without the need to deeply consider the text, or indeed understand the physical nature of the additive, the reader is immediately aware how these three vibratory ball milling reactions differ.

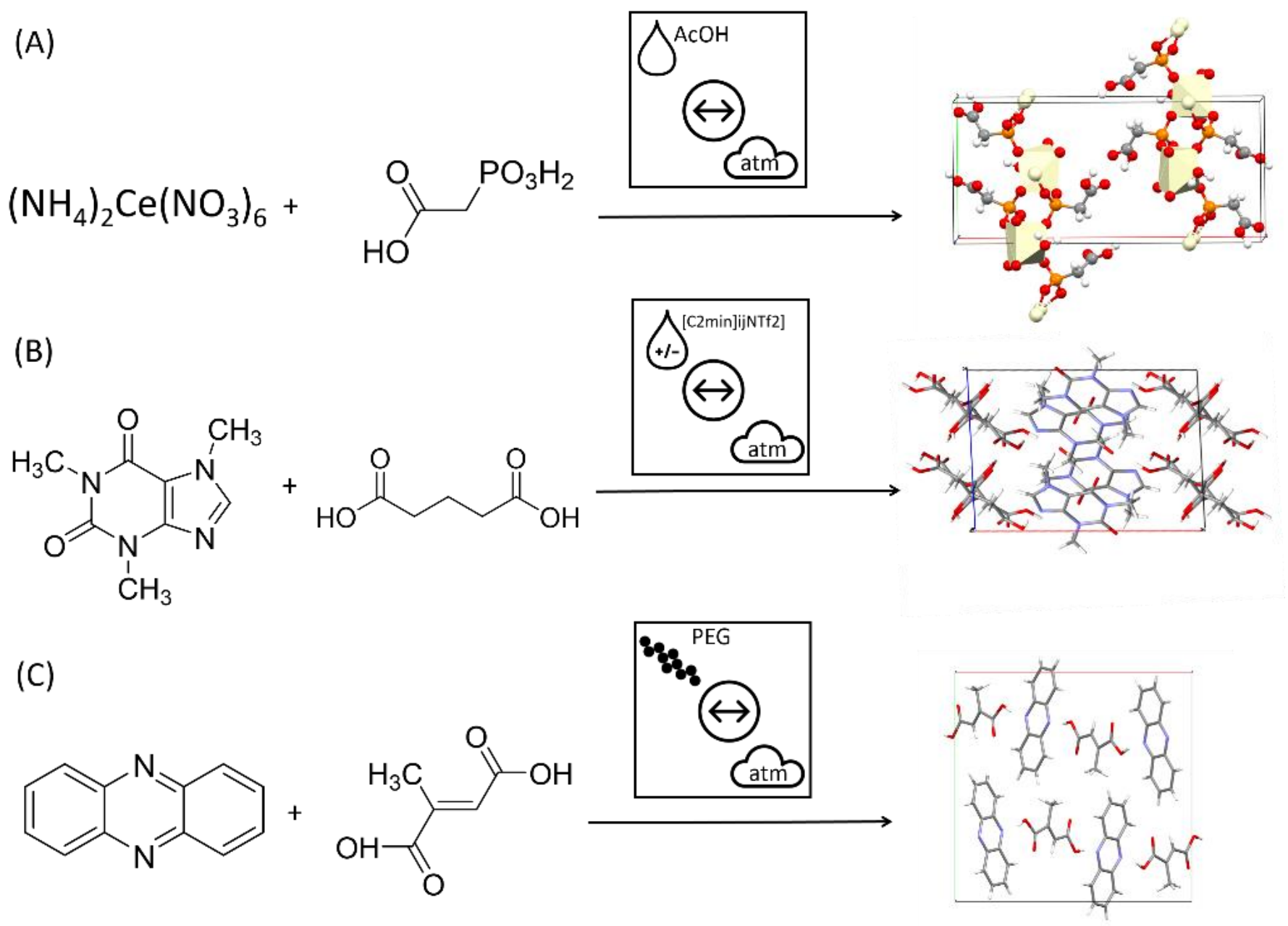

**Figure 10|** Using the proposed nomenclature to describe mechanochemical transformations with different types of additives. (A) An example of liquid assisted grinding formation of Ce based framework[239] (B) an example of ionic liquid assisted grinding for the formation of caffeine + glutaric acid cocrystals[240] (C) an example of polymer assisted grinding for the formation of phenazine + mesaconic acid cocrystals.[64]

Variation in temperature is also readily visible through use of the proposed nomenclature. This can be exemplified by the effects of cryo-temperatures on milling of $\epsilon$-chlorpropamide, Figure 11.[232] As previously discussed, the proposed nomenclature not only provides a thorough understanding of the mechanochemical conditions, but allows the reader immediate recognition of the role of temperature on this polymorphic transformation.

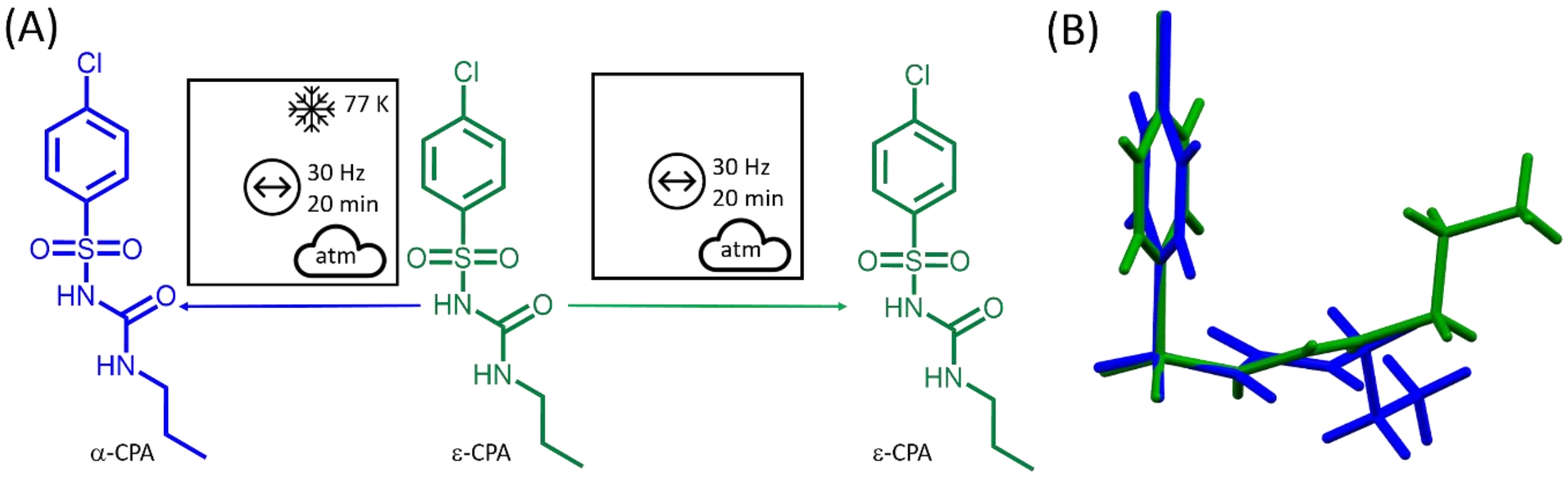

**Figure 11**: (A) Using the proposed nomenclature to describe mechanochemical transformations that differ according to temperature. (B) An illustration of the change in molecular conformations on cooling, which is reversible without a mechanical treatment, but is preserved (interlocked) after the molecular layers have been shifted on cryogrinding.

The need to specify explicitly the type of mechanical action within mechanochemical protocol is exemplified by the co-crystal formation of piroxicam and succinic acid, Figure 12.[197]The use of the triple milling ball symbol of Figure 12 would be wholly insufficient in such cases. The differing effects of shear and impact are instead captured explicitly within the proposed system, and a reaction scheme based upon this nomenclature allows immediate recognition of this important mechanochemical phenomenon.

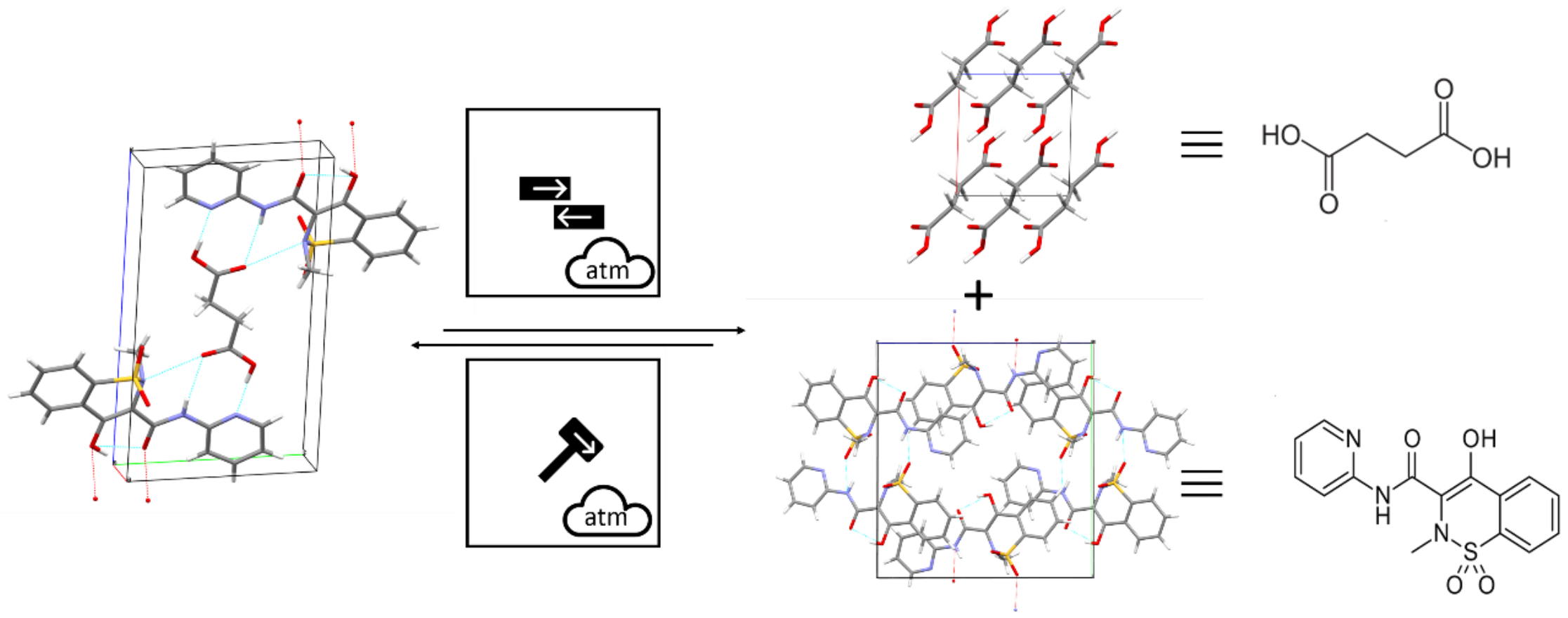

**Figure 12**: Using the proposed nomenclature to describe mechanochemical transformations that differ according to mechanical action. Shearing action leads to dissociation of the 2:1 piroxiam:succinic acid cocrystal phase into its coformers, whereas impact action on powder of the coformers leads to formation of the cocrystal.

We note that the pictographic representation is equally applicable to any reactions that occur as a result of mechanical action from across all aspects of chemical reactivity, Figure 13. Comparison of the pictographic representations for mechanochemical transformations of very different chemical species reveals immediately the diverse conditions required. It is readily apparent that inorganic compounds tend to be prepared using long duration planetary ball milling conditions under controlled atmospheres, whereas soft materials are prepared by gentler mechanical conditions and often benefit from the addition of liquid additives. Hence, not only does this pictographic representation allows rapid identification of experimental conditions being reported, but also offers a facile approach to identifying trends in experimental conditions successfully applied across the chemical and materials sciences. We therefore expect this clear and concise approach for representing mechanochemical transformations to facilitate new generalisations of mechanochemistry towards targeted and rapid design of new materials and molecular syntheses.

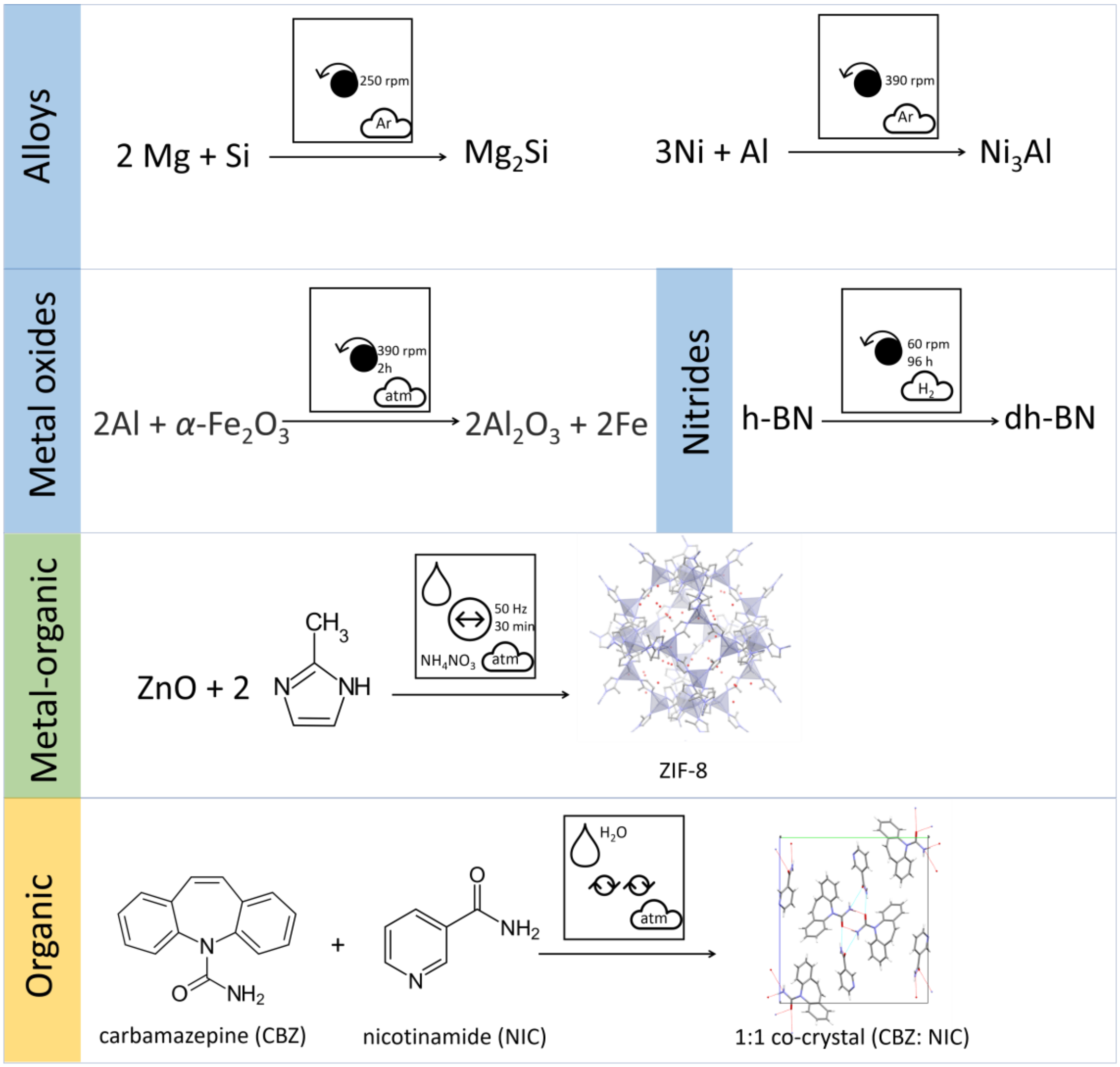

**Figure 13|** Pictographic representation of literature mechanochemical reactions for a diverse selection of chemical systems, including planetary milling of alloys,[241,242] metal oxides,[243] and nitrides[244] (e.g. hexagonal boron nitride h-BN converting to defect laden hexagonal boron nitride dh-BN), vibratory ball milling of metal organic frameworks,[245] and Resonant Acoustic Mixing cocrystal synthesis.[161]

## SUMMARY

The mechanochemistry community has expanded significantly in recent years. Throughout most of the 20th century, mechanochemistry was the focus of a smaller and relatively homogeneous community of predominantly solid-state scientists. Now, the community of mechanochemists has flourished, incorporating experts from a wide range of scientific backgrounds. To date, membership in the mechanochemistry field includes researchers from all branches of chemistry, physics, materials sciences, pharmaceutical sciences, biological sciences, and engineering. Many of the newcomers are themselves trained experts in solution or gas phase reactivity, thereby bringing with them many unique viewpoints on phenomena of chemical reactivity. The diversification of the mechanochemical community has triggered new, challenging and fundamentally important scientific questions and has led to mechanochemistry achieving more global impact than ever before.

Mechanochemistry today is understood to include much more than what is strictly defined by IUPAC as a "chemical reaction that is induced by the direct absorption of mechanical energy".[177] The term mechanochemistry has grown to include any transformation of that is observed during or after any type of mechanical treatment, regardless of the exact role of the mechanical action. Moreover, the term mechanochemistry is applied equally to transformations which occur upon stretching of single molecules, through to transformations within and between solids (including those which involve fluid intermediate states). Any transformation that is somehow facilitated by mechanical energy, or reactions that results from thermal- or photochemically induced stress and strain in a solid (the chemomechanochemical effect[246]) seem now to be denoted as 'mechanochemical'. Thermal or photochemical transformations in solids which have been mechanically pre-treated are similarly denoted as being 'mechanically activated', and hence also fall within the current paradigm of mechanochemistry.[246–248]

With this growing diversity of the community and the phenomena being explored comes a confusion of scientific languages on the scale of the Tower of Babel. Terminologies and jargon used by experts from one discipline are often misunderstood by experts from a different background. Similarly, much of the over a century's worth of research in mechanochemistry that is written in the scientific language of 20th century mechanochemical pioneers remains largely incomprehensible to many who enter the field. Moreover, many of these original works have been not digitalized, and are thus not easily accessible until recently. Advances in modern digital technology have made these precious papers and their translations available *via* online platforms, thereby allowing the global community to stand on the shoulders of the ancestral mechanochemical giants. Despite these digital advances, there is still a large miscommunication between the established and emerging mechanochemistry communities. Many well-documented phenomena are being unfortunately regularly re-discovered, with many new terminologies being coined to describe them.

Although in principle, science does not care how it is called, this has the knock-on effect of hindering how the scientific community can discuss, communicate, interpret, search the literature, and hence progress its collective understanding of the field. For example, many who accomplish an organic synthesis in a mechanical device do not realize that they in fact deal not only with a chemical transformations, but with a plethora of tribochemical phenomena. By considering only one aspect of the whole one risks to miss the elegance that nature has laid before us. For an elephant investigated in parts by blind men may be easily mistaken for a rope, a leaf, or a wall. In this same way, the strict isolation of tribo- and mechano-chemistry exists only in the minds of humans.

The growing interest in the mechanochemistry of organic compounds has revealed many new parameters which must be controlled to successfully achieve the reproducible mechanosynthesis of molecules and materials. These new parameters are of course in addition to those that were traditionally considered in tribochemistry and inorganic mechanochemistry. Parameters which are presently known to influence mechanochemical transformations include: the starting polymorph; the size and shape of particles; the type of mechanical action; the atmosphere under which the reaction has occurred; the presence and quantities of additives (solids, liquids, gases, polymers) even if they may not obviously participate in the reaction; the presence of catalysts, including those present as the materials of the milling bodies or reactors. Moreover, the specific parameters associated with the type of mechanical actions (e.g. revolutions per minute in twin screw extrusion or planetary ball milling, or frequency in ball milling, or the number, size and mass of the balls) are certainly important in defining the reaction. Yet, many such parameters are often overlooked, remain unreported, or are difficult to identify in literature reports.

Although efforts at unraveling the mechanistic aspects and driving forces of mechanochemical research are without doubt of central importance, effective communication of protocol and processes do not require such an understanding. It might not be possible to classify unambiguously a reaction as 'mechano-chemical', 'tribo-chemical' or 'mechanically facilitated thermochemistry' without this mechanistic understanding, but we can still impose clarity of communication when reporting our results. Following the successful example of the Crystallographic Information File (CIF), we demonstrate how adopting a standard format for reporting experimental conditions can help ensure that important parameters are both monitored and controlled. Such agreements on the type of information that needs to be presented and how it should be presented so that it is clear to all on first glance, are crucial to unify the community and drive fundamental developments in the field of mechanochemistry.

We believe this *opinion* piece will spark timely and productive discussion across the ever growing and diversifying mechanochemistry community. All symbols displayed in this text are available free of charge for those who wish to use them. Additional pictograms and information will certainly become important as new features of control over mechanochemical reactions emerge, and we as a community must be ready to adapt our nomenclature and standard of practice to accommodate this.

As mechanochemistry becomes an increasingly integral part of efforts to develop environmentally benign strategies for chemical and industrial processes, it is timely to make efforts to unite the dynamic community behind common concepts and definitions in scientific language, and terminology. Only in doing so can we hope to avoid the potential downfall of the 'Tower of Babel' and construct a coherent and robust 'Tower of Mechanochemistry'. It is only by uniting the community that we can collectively obtain knowledge as, in the words of N. Copernicus, 'to know that we know what we know, and that we do not know what we do not know, that is true knowledge.'

**Acknowledgements**

As the founding president of the IMA (VVB) and as members of the IMA and COST Action CA18112, the authors acknowledge support of the ongoing efforts of both associations to develop a global community of mechanochemists. The authors are grateful to the many significant and seminal contributions to the field of mechanochemistry by many remarkable scientists, upon whose shoulders the current field stands. EVB acknowledges the financial support from RFBR (grant 19-29-12026-мк) when working on this paper.

**Notes**

A library of the pictograms used in this study are available free of charge from https://opus4.kobv.de/opus4-bam/frontdoor/index/index/docId/52329